\documentclass[twocolumn,10pt]{article}
\usepackage[noblocks]{authblk}
\usepackage{tabularx}
\usepackage{graphicx}
\usepackage{multicol}
\usepackage{bm}
\usepackage{units}
\usepackage{amsmath}
\usepackage{amssymb}
\usepackage{color}
\usepackage{ulem}
\usepackage{comment}
\usepackage[switch]{lineno} 
\usepackage[squaren,Gray]{SIunits}
\usepackage[dvipsnames]{xcolor}
\usepackage[utf8]{inputenc} 
\usepackage{float}
\usepackage{mcite}
\usepackage{hyperref}
\usepackage{array}
\newcolumntype{L}[1]{>{\raggedright\let\newline\\\arraybackslash\hspace{0pt}}m{#1}}
\newcolumntype{C}[1]{>{\centering\let\newline\\\arraybackslash\hspace{0pt}}m{#1}}
\newcolumntype{R}[1]{>{\raggedleft\let\newline\\\arraybackslash\hspace{0pt}}m{#1}}
\let\OLDthebibliography\thebibliography
\renewcommand\thebibliography[1]{
  \OLDthebibliography{#1}
  \setlength{\parskip}{0pt}
  \setlength{\itemsep}{0pt plus 0.3ex}
}
\makeatletter
\def\@maketitle{%
  \newpage
  \null
  \vskip 4.em%
  \begin{center}%
  \let \footnote \thanks
    {\huge \bfseries \@title \par}%
    \vskip 0.0em%
    {\normalsize
      \lineskip 0.em%
      \begin{tabular}[t]{c}%
        \@author
      \end{tabular}\par}%
    \vskip 0.em%
    {\normalsize \@date}%
  \end{center}%
  \par
  \vskip 1.3em}
\makeatother
\newcommand{\APC}{APC, CNRS/IN2P3, CEA/IRFU, Observatoire de Paris, Sorbonne Paris Cit\'{e} University, 10 rue Alice Domon et L\'{e}onie Duquet, 75205 Paris Cedex 13, France}
\newcommand{\MilanoUni}{Dipartimento di Fisica, Universit\`{a} di Milano-Bicocca, I-20126 Milano, Italy}
\newcommand{\MilanoINFN}{INFN, Sezione di Milano-Bicocca, I-20126 Milano, Italy}
\newcommand{\CBPF}{Centro Brasileiro de Pesquisas F\'{i}sicas (CBPF), Rua Xavier Sigaud 150, Rio de Janeiro, RJ, 22290-180, Brazil}
\newcommand{\CENBG}{Universit\'e de Bordeaux, CNRS, CENBG-IN2P3, F-33170 Gradignan, France}
\newcommand{\CIEMAT}{CIEMAT, Centro de Investigaciones Energ\'{e}ticas, Medioambientales y Tecnol\'{o}gicas,  Av. Complutense 40, E-28040 Madrid, Spain}
\newcommand{\Chooz}{LNCA Underground Laboratory, CNRS/IN2P3 - CEA, Chooz Nuclear Reactor, 08600 Chooz, France}
\newcommand{\CPPM}{Aix Marseille Univ, CNRS/IN2P3, CPPM, Marseille, France}
\newcommand{\FerraraUni}{Department of Physics and Earth Sciences, University of Ferrara, Via Saragat 1, 44122 Ferrara, Italy}
\newcommand{\FerraraINFN}{INFN, Ferrara Section, Via Saragat 1, 44122 Ferrara, Italy}
\newcommand{\UCI}{Department of Physics and Astronomy, University of California at Irvine, 4129 Frederick Reines Hall, Irvine, California 92697, USA}
\newcommand{\IJCLab}{IJCLab, CNRS/IN2P3, Universit\'e Paris-Saclay, Universit\'e de Paris, 15 rue Georges Cl\'emenceau, 91400, Orsay, France}
\newcommand{\Londrina}{Departamento de F\'isica, Universidade Estadual de Londrina, Rodovia Celso Garcia Cid, PR 445 Km 380, Campus Universit\'ario Cx. Postal 10.011, CEP 86.057-970, Londrina -- PR, Brazil}
\newcommand{\MPIK}{Max-Planck-Institut f\"{u}r Kernphysik, Saupfercheckweg 1, 69117 Heidelberg, Germany}
\newcommand{\Padovauni}{Dipartimento di Fisica e Astronomia, Universit\`{a} di Padova, Via Marzolo 8, I-35131 Padova, Italy}
\newcommand{\Padova}{INFN, Sezione di Padova, via Marzolo 8, I-35131 Padova, Italy}
\newcommand{\PUCR}{Department of Physics, Pontif\'icia Universidade Cat\'olica do Rio de Janeiro, C.P. 38097, 22451-900, Rio de Janeiro, RJ, Brazil}
\newcommand{\PUCSPhys}{Pontificia Universidad Cat\'olica de Chile, Avda. Vicuna Mackenna 4860, Macul, Santiago, Chile}
\newcommand{\Queens}{Department of Physics, Engineering Physics \& Astronomy, Queen's University, Kingston, Ontario K7L3N6, Canada}
\newcommand{\RCNS}{RCNS, Tohoku University, 6-3 AzaAoba, Aramaki, Aoba-ku, 980-8578, Sendai, Japan}
\newcommand{\SUBA}{SUBATECH, CNRS/IN2P3, Universit\'{e} de Nantes, IMT-Atlantique, 44307 Nantes, France}
\newcommand{\Sussex}{Department of Physics and Astronomy, University of Sussex, Falmer, Brighton BN1 9QH, United Kingdom}
\newcommand{\TorinoINFN}{INFN, Sezione di Torino, Via P. Giuria 1, I-10125 Torino, Italy}

\newcommand{\n}{{$\nu$}}
\newcommand{\RC}{Cowan, Reines~et~al.}

\begin{document}
\normalem 
%
%
\author[1,2,3]{A.~Cabrera}
\author[4]{A.~Abusleme}
\author[5]{J.~dos~Anjos}
\author[6,7]{T.~J.~C.~Bezerra}
\author[2]{M.~Bongrand} 
\author[2]{C.~Bourgeois}
\author[2]{D.~Breton}
\author[8]{C.~Buck} 
\author[9]{J.~Busto} 
\author[10]{E.~Calvo}
\author[11]{E.~Chauveau}
\author[12]{M.~Chen} 
\author[13]{P.~Chimenti} 
\author[14]{F.~Dal~Corso}
\author[13]{G.~De~Conto}
\author[14]{S.~Dusini} 
\author[15,16]{G.~Fiorentini}
\author[13]{C.~Frigerio~Martins}
\author[1]{A.~Givaudan}
\author[17,18]{P.~Govoni}
\author[8]{B.~Gramlich}
\author[1,2,19]{M.~Grassi}
\author[1,2]{Y.~Han}
\author[7]{J.~Hartnell} 
\author[9]{C.~Hugon}
\author[10]{S.~Jiménez}
\author[1]{H.~de~Kerret\thanks{Deceased.}}
\author[2]{A.~Le~Nev\'e}
\author[2]{P.~Loaiza}
\author[2]{J.~Maalmi}
\author[15,16]{F.~Mantovani} 
\author[2]{L.~Manzanillas}
\author[11]{C.~Marquet} 
\author[6]{J.~Martino}
\author[2,10]{D.~Navas-Nicol\'as}
\author[20]{H.~Nunokawa} 
\author[1]{M.~Obolensky}
\author[21]{J.~P.~Ochoa-Ricoux} 
\author[22]{G.~Ortona}
\author[10]{C.~Palomares} 
\author[20]{F.~Pessina}
\author[11]{A.~Pin}
\author[7]{J.~C.~C.~Porter} 
\author[11]{M.~S.~Pravikoff}
\author[11]{M.~Roche}
\author[21]{B.~Roskovec}
\author[2]{N.~Roy}
\author[1]{C.~Santos}
\author[8]{S.~Schoppmann}
\author[15,16]{A.~Serafini}
\author[2]{L.~Simard}
\author[17]{M.~Sisti} 
\author[14]{L.~Stanco}
\author[15,16]{V.~Strati}
\author[6]{J.-S.~Stutzmann}
\author[1,23]{F.~Suekane}
\author[10]{A.~Verdugo}
\author[6]{B.~Viaud}
\author[1]{C.~Volpe}
\author[1]{C.~Vrignon}
\author[1]{S.~Wagner}
\author[6]{F.~Yermia}
%
%
%
\affil[1]{\APC}
\affil[2]{\IJCLab}
\affil[3]{\Chooz}
\affil[4]{\PUCSPhys}
\affil[5]{\CBPF}
\affil[6]{\SUBA}
\affil[7]{\Sussex}
\affil[8]{\MPIK}
\affil[9]{\CPPM}
\affil[10]{\CIEMAT}
\affil[11]{\CENBG}
\affil[12]{\Queens}
\affil[13]{\Londrina}
\affil[14]{\Padova}
\affil[15]{\FerraraINFN}
\affil[16]{\FerraraUni}
\affil[17]{\MilanoINFN}
\affil[18]{\MilanoUni}
\affil[19]{\Padovauni}
\affil[20]{\PUCR}
\affil[21]{\UCI}
\affil[22]{\TorinoINFN}
\affil[23]{\RCNS}
%
%

\title{\bf Neutrino Physics with an Opaque Detector}
\date{(LiquidO Consortium)\thanks{E-mail: LiquidO-Contact-L@in2p3.fr} \\ \today}

\maketitle

{
\bf 
\noindent
In 1956 Reines~\&~Cowan discovered the neutrino using a liquid scintillator detector. The neutrinos interacted with the scintillator, producing light that propagated across transparent volumes to surrounding photo-sensors. This approach has remained one of the most widespread and successful neutrino detection technologies used since. This article introduces a concept that breaks with the conventional paradigm of transparency by confining and collecting light near its creation point with an opaque scintillator and a dense array of optical fibres. This technique, called LiquidO, can provide high-resolution imaging to enable efficient identification of individual particles event-by-event. A natural affinity for adding dopants at high concentrations is provided by the use of an opaque medium. With these and other capabilities, the potential of our detector concept to unlock opportunities in neutrino physics is presented here, alongside the results of the first experimental validation.
}
\vspace{0.cm}

\section*{Introduction}
\noindent The discovery of the neutrino (\n) in the fifties~\cite{Ref_NuDiscovery} revolutionised particle physics not only by establishing the existence of this elusive particle, but also by laying the foundations for a technology used in many subsequent breakthroughs. The liquid scintillator detector (LSD) developed by \RC\ for \n~detection exploited a well established radiation detection technique at the time, whereby molecular electrons are excited by the passage of charged particles produced by \n~interactions and then emit light upon de-excitation~\cite{Ref_LS}. This light is detected by sensitive photon detectors, typically photo-multiplier tubes (PMTs)~\cite{Ref_PMT}, that surround the scintillator volume and are often located many metres from the interaction point. \RC~relied on the inverse-$\beta$ decay (IBD) reaction, given by $\bar\nu_e + p \to e^+ + n$, that yields two clear signals: the prompt energy deposition of the $e^+$ (including annihilation $\gamma$'s) followed by the nuclear capture signal of the $n$ after thermalisation. 
The close time and space coincidence between these two was exploited as the primary handle to separate the signal from the background~(BG). The simplicity and power of this technique, enabled in great part by the abundant light produced by the scintillators, has allowed LSDs to dominate several areas of neutrino physics, particularly at the lower part of the MeV energy scale.

Despite their many advantages, LSDs have limitations. The propagation of light through the scintillator itself makes transparency an essential requirement for efficient light collection, potentially limiting the size of the detector volume. Given the extremely small probability for neutrinos to interact with matter, achieving larger detectors has in fact been a standing challenge throughout the history of \n~physics. 
LSDs have gone from a few hundred kilograms, at the time of \RC, to today's 20~kilotons with JUNO~\cite{Ref_JUNO}, where a record setting mean attenuation length of greater than 20~m is foreseen. The need for transparency has also set tight constraints on the type and concentration of elements that can be loaded into the scintillator. The physics goals of certain experiments call for detector doping~\cite{Ref_DopingGeneral}, where an element other than the scintillator's native H and C is added to enhance detection capabilities or to search for rare processes. The discovery of the neutrino itself involved doping the detector with $^{113}$Cd~\cite{Ref_RC-Loading} to increase the energy released on $n$ capture and thus further reduce the BG. However, to this day, doping in LSDs has been limited at high concentrations by transparency and stability constraints.

LSDs typically have a rather poor event-by-event topological discrimination power. It is essentially impossible to distinguish an individual $e^+$ from an $e^-$ or a $\gamma$ below 10~MeV, and it is even very difficult to tell whether one or several events have occurred simultaneously. The primary approach to deal with these limitations has been to segment the detector. 
Here the main volume is subdivided into optically-decoupled compartments, so instead of a single monolithic volume a granular one is used. 
This allows recovery of topological information from each neutrino interaction, i.e. images of the space and time pattern of an event, and hence enhances a detector's event identification capability. 
This technique has been successfully used with GeV-scale neutrinos, where the large physical extent of the events allows imaging of the final state particles using segmentation of a few centimetres. The largest example today is the 14~kiloton NOvA detector~\cite{Ref_NOvA_1,Ref_NOvA_2}, where the resulting images of the neutrino events are crucial for BG rejection and identification of different types of neutrino interactions. The situation is more difficult with MeV-scale neutrino interactions given the smaller physical extent of the resulting energy depositions, although there can still be advantages to segmenting. For instance, the coarsely segmented detector introduced by \RC~\cite{Ref_NuDiscovery} exploited the unique anti-matter annihilation pattern of the $e^+$, producing two back-to-back mono-energetic $\gamma$'s, as an aid to event identification. It is difficult however to segment finely enough to resolve the full topological information of individual events at these low energies without introducing certain disadvantages, such as dead material, radioactivity, cost, etc.

Since the 1950s numerous discoveries have been made using \n's from reactors, the sun, the interactions of cosmic rays in the atmosphere, accelerator produced beams, one supernova explosion (SN1987A), the earth and other astrophysical sources~\cite{Ref_PDGnu}, and LSDs have played key roles in many of them.
Despite the remarkable progress, the limitations of today's detector technology constrain our ability to probe the \n~beyond our current knowledge and to use it in further exploring the universe. 

This article presents a technique for neutrino detection, called LiquidO, that uses an opaque scintillator and a lattice of optical fibres to confine and collect light near its creation point. Extensive studies with simulations have been performed showing that our approach possesses many of the strengths of the existing technology while also giving rise to other capabilities, such as high-resolution imaging and a more natural affinity for doping. The principles of the technique have been demonstrated with a small experimental setup. The result is a detector concept with the potential to break ground in various frontiers of neutrino physics, some of which have remained elusive for decades.

\section*{Results and Discussion}
\begin{figure*}[!ht]
	\centering
	\includegraphics[scale=0.94]{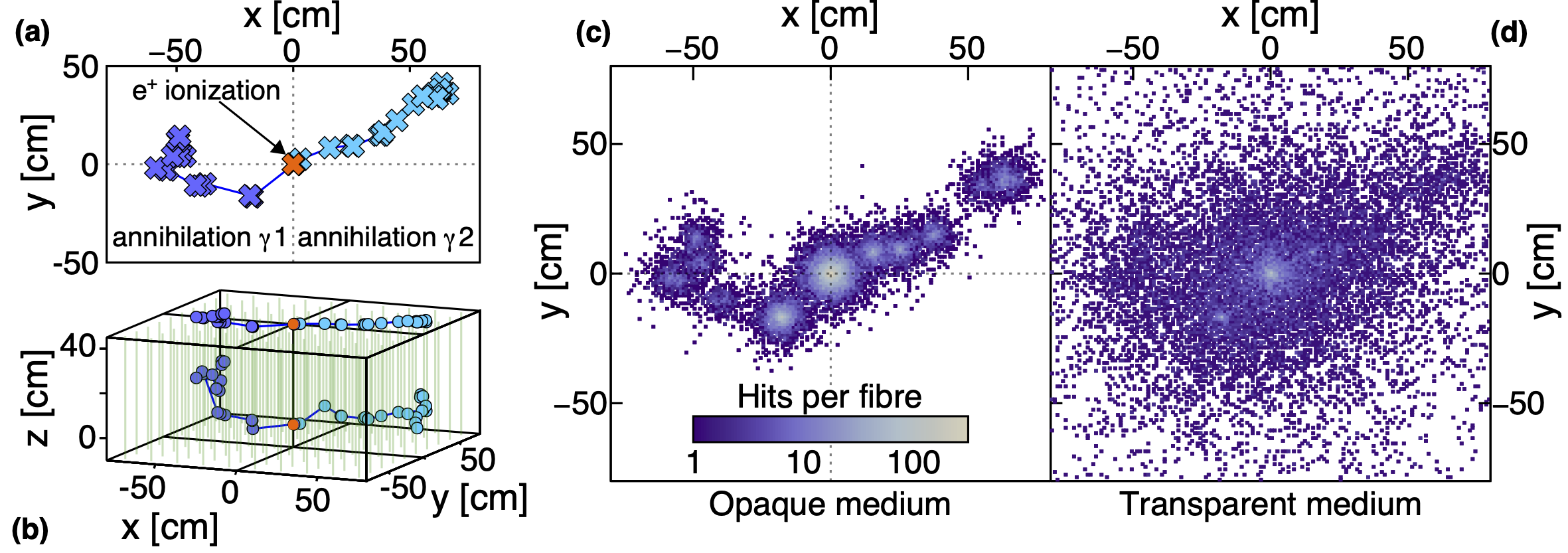}
	\caption{ \small 
	\textbf{Detection Principle and Imaging Capability.} 
	The energy depositions in a LiquidO detector from a simulated $e^+$ with 1~MeV of kinetic energy plus the associated annihilation gammas are used to illustrate the detection principle and imaging capability. {\bf a} shows a two-dimensional $x$-$y$ projection of the energy depositions in the scintillator, while {\bf b} shows the full three-dimensional extent. The green lines running in parallel with the $z$-axis represent the fibres in the simplest configuration of the detector geometry.
	The energy deposition of the $e^+$ is shown by the orange point and the Compton scatters of the 0.51~MeV back-to-back annihilation $\gamma$'s by light and dark blues. 
	One of the $\gamma$'s (dark blue) turns upwards to run approximately parallel with the fibres and this is reflected by its shorter extent in the $x$-$y$ projection. 
	The simulation of the light hitting each fibre in a 1-cm-pitch lattice is shown in two scenarios: 
	on {\bf c} an opaque scintillator with a 5~mm scattering length is simulated, whereas on {\bf d} the scintillator is transparent. The colour of each point represents the number of photons hitting a fibre at that $x$-$y$ location. 
	With a transparent medium the resulting image from the fibre array is almost completely  washed out. In stark contrast, the light confinement around each energy deposition with the opaque scintillator allows preservation of the event's precious topological information and the formation of a high-resolution image.
	}
	\label{Fig_1}
\end{figure*}

\subsection*{\label{sec:detection}Detection Principle}
\noindent Our detection technique is based on using an opaque scintillator. 
Opacity can be achieved in two ways, through light scattering and/or absorption. 
The LiquidO approach relies on a short scattering length and an intermediate or long absorption length, producing a scintillator that is milky and translucent in appearance.
Photons from the scintillator undergo a random walk about their origin, giving rise to stochastic light confinement. 
While the path of each photon is stochastic, the integral effect is the confinement of the light to a sphere around each ionisation point, resulting in the production of so-called light-balls.
This is the principle at the heart of the LiquidO technique. 

Scintillators used in modern neutrino experiments typically have scattering lengths of up to tens of metres. Reducing the scattering length down to the scale of millimetres causes the light to be confined to a volume that is much smaller than the typical physical extent of, for example, a 1~MeV $\gamma$-ray event whose energy is lost via Compton scattering. 
To extract the light a lattice of wavelength-shifting fibres runs through the scintillator. 
With a lattice spacing on the scale of a centimetre, prompt and efficient light collection can be achieved and absorption losses minimised.
Many configurations of the fibre lattice are possible, and in principle fibres could run in all three orthogonal directions. In practice, fibres running in one or two directions might suffice. The difference in time of the light detection at the two ends of each fibre can be used as a measure of the event position along the length of each fibre.

The detection principle using the simplest configuration, where the fibres all run along the $z$-axis, is illustrated in Fig.~\ref{Fig_1}. The energy depositions in three-dimensional space from a simulated $e^+$ with 1~MeV of kinetic energy are shown in Fig.~\ref{Fig_1}b, while Fig.~\ref{Fig_1}a shows the two-dimensional $x$-$y$ projection.
A simulation of the light propagation is shown in Fig.~\ref{Fig_1}c, d, where the colour of each point represents the number of photons hitting a fibre. An opaque scintillator is simulated in Fig.~\ref{Fig_1}c and in Fig.~\ref{Fig_1}d the scintillator is transparent. 
The formation of the light-ball around the position of each Compton electron can be seen clearly for the opaque scintillator, whereas that pattern is almost completely washed out in the transparent case. A finely segmented detector would be required to measure this topology using transparent scintillator. In contrast, the LiquidO technique effectively self-segments due to stochastic light confinement. This eliminates the need to add dead material (with associated potential radioactivity) to achieve segmentation and therefore substantially reduces the cost and complexity of producing scintillator detectors capable of high-resolution imaging.

The development of the LiquidO approach builds on much of the well-established technology of scintillator detectors, including photo-sensors, wavelength-shifting fibres and the organic scintillator materials themselves. Modern Si-based photo-sensors~(SiPM)~\cite{Ref_SiPM_1, Ref_SiPM_2}, with high quantum efficiencies of 50\% and time resolutions as fast as 100~ps per photon detected~\cite{Ref_TimePerSiPM_LAL}, present themselves as an excellent option for LiquidO. From our simulations, it is estimated that more than 90\% (60\%) of the light will hit the fibres of a 1-cm-pitch lattice in a scintillator with an absorption length of 5~m (1~m). Compared to the tens of metre long absorption lengths necessary for the largest LSD based experiments, this represents a substantial reduction in the requirements. 
The same techniques used to purify scintillators that have been successfully used in experiments like Borexino~\cite{Ref_BX} could be used in LiquidO as needed. The potential contamination introduced by the presence of fibres in the scintillator volume is mitigated by the fact that they amount to less than 1\% of the detector mass fraction and that excellent levels of fibre radiopurity have been achieved~\cite{Ref_GERDAFibre}. For physics measurements requiring extremely low backgrounds from natural radioactivity (at energies below 3~MeV), further improvements in fibre radiopurity may be necessary.

With a typical organic scintillator light yield of about 10~photons per keV~\cite{Ref_LS}, a 5~m absorption length~\cite{Ref_NOWASH}, and a wavelength-shifting fibre acceptance of about 10\% (the main loss in detection~\cite{Ref_FibreAcceptance}), the number of detected photons is estimated to be a maximum of around 400~photo-electrons per MeV for a small 1~cm-pitch lattice detector. When scaling to larger sizes this amount will reduce due to the several-metre attenuation lengths typical of wavelength-shifting fibres. The optimisation of the light collection can include consideration of elongated geometries, modularisation, and/or  double-ended readout, whose cost can be strongly mitigated by multiplexing. An exciting aspect of our detector concept is that it makes scintillating materials that are naturally opaque ideal, opening up a whole landscape of substances to explore. Known scintillators with substantially higher light output present promising avenues of research, alongside the possibility of finding new materials that have simply not been carefully studied yet due to their poor transparency.

The energy resolution of our 1~cm-pitch detector with a light yield of 400~photo-electrons per MeV is estimated from simulations to be $5\%/\sqrt{E {\rm (MeV)}}$ as expected where Gaussian statistics dominate. The position-dependent response of our baseline detector is very uniform after attenuation in the fibres is calibrated out. It varies by less than 1\% across more than 95\% of the volume and has a negligible effect on the energy resolution.

\begin{figure}[!ht]
	\centering 
	\includegraphics[scale=0.47]{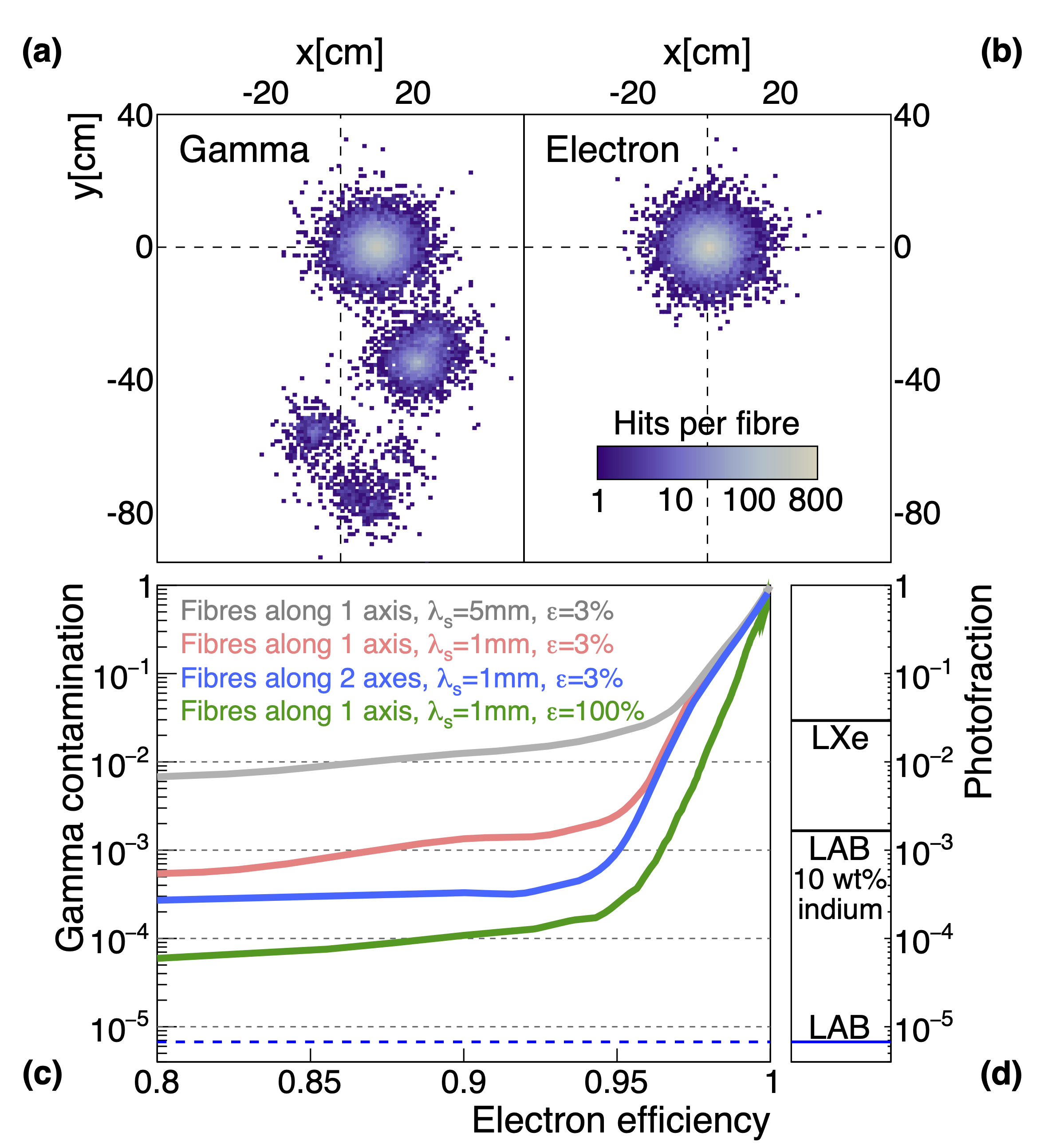}
	\caption{ \small 
	\textbf{Discrimination of Electrons from Gammas.}
	An image of a $\gamma$ ({\bf a}) and an $e^-$ ({\bf b}) with 2~MeV of kinetic energy simulated using the default LiquidO detector configuration with fibres arranged in a 1-cm-pitch lattice running along the $z$-axis. The spatially dispersed Compton-scattering pattern of the $\gamma$ clearly sets it apart from the $e^-$. 
	{\bf c} shows the probability of misidentifying a $\gamma$ as an $e^-$ vs.\ the efficiency of selecting $e^-$'s estimated with a simple reconstruction. 
	The scintillator is assumed to have a conventional light yield and a mean scattering length $\lambda_s$ of either 1~mm or 5~mm, which is well-matched to the 1~cm fibre pitch. The photon detection efficiency $\varepsilon$ of $3\%$ accounts for all losses of light at the various stages and is dominated by the fibre trapping efficiency (around 10\%) and the Si-based photo-sensor (SiPM) quantum efficiency (around 50\%). 
	The grey curve  shows the probability of misidentifying a 2~MeV $\gamma$ as an $e^-$ is estimated to be at the $10^{-2}$ level with an efficiency of 87\% for $\lambda_s$=5~mm. The red curve illustrates how the $\gamma$ contamination decreases by an order of magnitude for the same efficiency when the light is more tightly confined ($\lambda_s$=1~mm).
	The blue curve shows the improvement that can be obtained when fibres with the same $1$~cm pitch run along 2 orthogonal axes instead of a single one. Finally, the green curve illustrates how much hypothetical improvement could be obtained in the limit of 100\% efficiency or equivalently 30$\times$ more light, some of which might be achieved through novel scintillators and/or improved photon detection efficiency. No timing information has been used and more sophisticated spatial reconstruction techniques could likely improve the $e^-$ vs. $\gamma$ separation further.
	A photofraction scale is shown on {\bf d}, which quantifies the fraction of $\gamma$'s that interact via the photo-electric effect and thus provides a floor to detector performance. Linear-Alkylbenzene (LAB) has a low photofraction of $6\times10^{-6}$, compared to 2.9\% for liquid xenon and 0.17\% for LAB doped with 10\% indium by weight.	
	}
    \label{Fig_2}
\end{figure}

\subsection*{\label{sec:capabilities}Imaging and Particle Identification}
The intrinsic high-resolution imaging capability of our technique is one of its main advantages. 
Using the information on the quantity of light collected from each fibre, the position of a point-like energy deposition can be reconstructed to within a few millimetres in the transverse direction to the fibres. This level of precision enables discrimination between point-like events such as MeV-scale $e^-$'s and events with spatially dispersed energy depositions such as $e^+$'s and $\gamma$'s.
This is illustrated in Fig.~\ref{Fig_2}a, b, where the topology of a $\gamma$ and an $e^-$ with 2~MeV of kinetic energy can be compared. 
The $e^-$ deposits all its energy within a centimetre whereas the $\gamma$ Compton scatters over many tens of centimetres. 
The discrimination power of LiquidO is quantified in Fig.~\ref{Fig_2}c where the probability that a $\gamma$ is misidentified as an $e^-$ is shown versus the $e^-$ selection efficiency.
A simple reconstruction algorithm quantifying the spatial spread of the hit fibres is used for these studies.
The results indicate that 2~MeV $e^-$'s can be feasibly distinguished from $\gamma$'s with a contamination factor better than $10^{-2}$, which is unprecedented for LSDs at these energies. Similarly, the topology of a $e^+$ annihilation event, with its back-to-back $\gamma$'s as illustrated in Fig.~\ref{Fig_1}c, stands in stark contrast to the point-like energy deposition of an $e^-$. 
Charged particles with enough kinetic energy to travel several cm or more in the detector will produce sequences of point-like energy depositions. Such track-like signatures would arise from, for example, muons, allowing their path through the detector to be precisely reconstructed. 
Track-like patterns would also be formed from many other particle interactions such as $\bar\nu_e$ charged current~(CC) above about 10~MeV and $\nu_\mu$~CC events at higher energies above the $\mu$ production threshold. In this way, LiquidO combines some of the advantages of tracking detectors with those of LSDs.

The timing information of the light pulses coming from each fibre is  expected to further enhance the particle identification capabilities of LiquidO. Our simulations show that $e^+$'s and $\gamma$'s have distinct energy-flow patterns, in that the $e^+$ event typically develops outwards from a central light-ball while the $\gamma$ consists of several light-balls forming in sequence. Work is ongoing to quantify the ability of LiquidO to perform dynamic imaging of energy depositions in time and the consequent improvement over the static imaging used in Fig.~\ref{Fig_2}a, b. 
If successful, this could allow single-$e^+$ events to be efficiently identified below 3~MeV, where most gamma backgrounds from natural radioactivity lie. Above this energy, we expect the timing information would typically be much less important and that the static images alone are likely to enable single-$e^+$ identification.

The particle identification capability of our technique builds on the low density of organic scintillator, typically 0.9~g~cm$^{-3}$, and its high fraction of hydrogen with H-to-C ratios typically in the range 2-3.
Its low average atomic number favours a long radiation length, around 0.5~m, a minimal photo-electric effect and energy losses by bremsstrahlung that do not start to dominate until $e^-$'s have an energy of around 100~MeV.\
The extremely low cross-section for the photo-electric effect in scintillator, such as Linear-Alkylbenzene (LAB)~\cite{Ref_LAB}, means that an MeV-scale $\gamma$ is highly unlikely to interact that way. 
On the rare occasion that this does happen an $e^-$ of the same energy is produced, which sets a limit to the level at which $e^-$'s and $\gamma$'s can be distinguished.
In scintillators, the fraction of 2~MeV $\gamma$'s that interact via the photo-electric effect (the photofraction) is only $6\times10^{-6}$ whereas in a heavy liquid such as xenon it is 2.9\%. 
Doping a scintillator with a metal causes the photofraction to increase. For example, if indium at 10\% by mass is used the photofraction rises to 0.17\%. 
These numbers are illustrated in Fig.~\ref{Fig_2}d, allowing comparison of the probability of the event reconstruction misidentifying a $\gamma$ as an $e^-$ with the floor to performance set by the detector material.

\subsection*{\label{sec:doping}Elemental Doping}
A particularly promising avenue for exploiting the LiquidO approach is where doping of the scintillator opens up the possibility of new physics measurements.
One of the major challenges usually associated with doping LSDs is maintaining the optical properties, including transparency, while achieving the desired concentration of the dopant.
In contrast, our technique actually requires opacity to confine the light and therefore allows for consideration of more possibilities, be it to load new materials or to achieve higher levels of doping. 
Examples of what can be achieved with a doped scintillator are wide and varied. 
The original \RC\ experiment used cadmium to increase the neutron capture cross-section and the LENS experiment concept involved using an indium-doped liquid scintillator~\cite{Ref_DopingLENS_1, Ref_DopingLENS_2, Ref_DopingLENS_3}. Several neutrino-less double beta decay experiments use or propose doped scintillators~\cite{Ref_KLZ,Ref_SNO+,Ref_SNO+_2,Ref_JUNODBD,Ref_Talk_NOW} as the way forward to realise higher isotopic masses.
The strong precedent set by LENS with indium suggests that loading at more than 10\% for neutrino-less double beta decay searches is a reachable objective.

\subsection*{\label{sec:demonstration}First Experimental Proof of Principle}

\begin{figure*}[!ht]
	\centering 
	\includegraphics[scale=0.65]{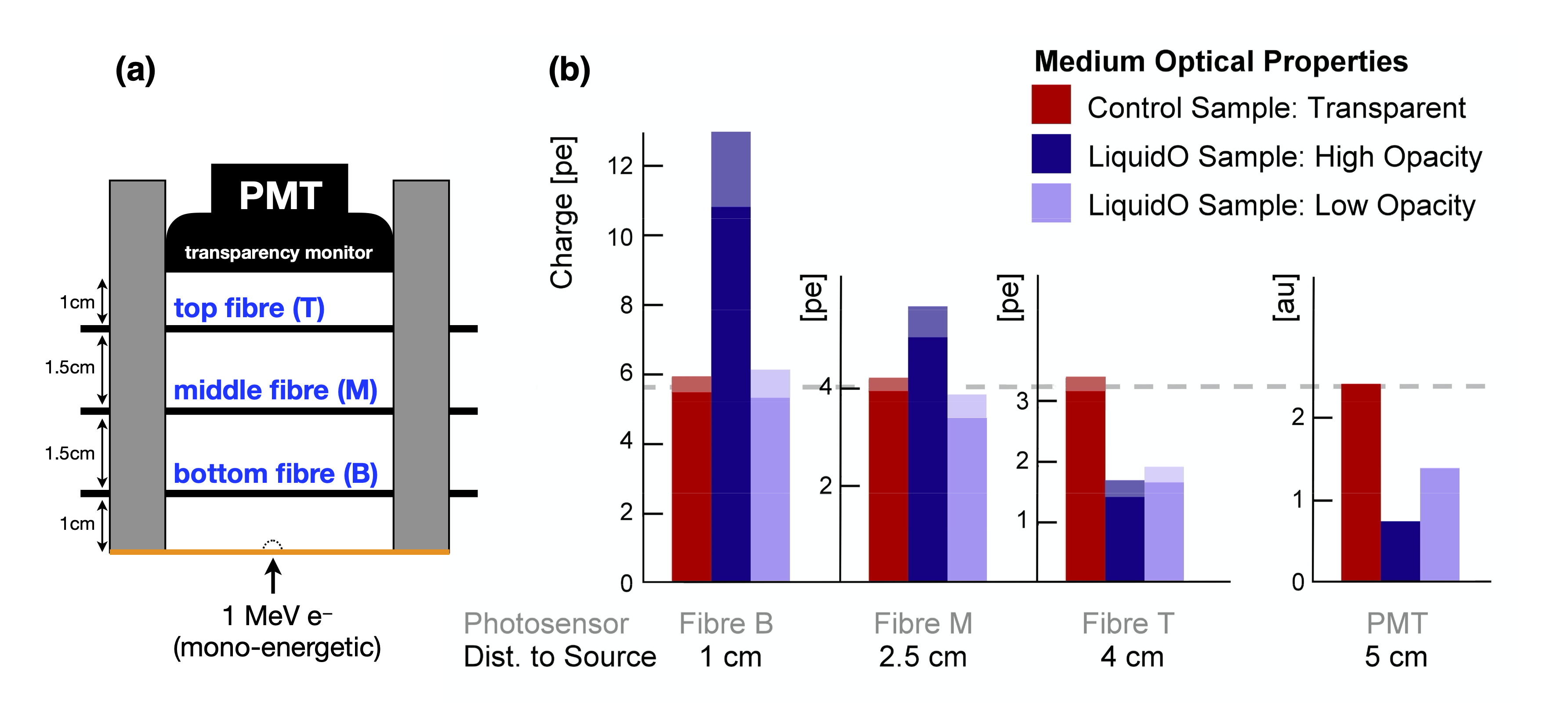}
	\caption{ \small 
	{\bf Experimental Proof of Principle.}
	{\bf a} Diagram of the small prototype detector built to make the first experimental demonstration of our opaque scintillator detection concept.
	The detector consists of a small cylindrical vessel with internally reflecting surfaces including a 25~$\micro$m aluminised Kapton sheet at the bottom.
	Light is collected with three fibres ($\phi$=1~mm) running diametrically at different heights. 
	A 3" photo-multiplier tube (PMT) is placed at the top and serves as a transparency monitor. 
	Mono-energetic 1~MeV $e^-$'s impinge from the bottom and make point-like energy depositions inside the detector as shown by the dashed semi-circular line.
	{\bf b} The data collected with three scintillators made from the same base of Linear-Alkylbenzene (LAB) plus a PPO wavelength-shifter at 2~g~l$^{-1}$ are shown with measurement uncertainties illustrated by the pale regions at the top of each bar. These uncertainties correspond to the standard deviation of up to 10 measurements for each sample.
	The high (dark blue) and low (light blue) opacity formulations are obtained by mixing in a paraffin polymer at 10\% and setting the temperature at 12$\degree$C and 26$\degree$C, respectively. 
	The measurements from the prototype obtained with the opaque formulations are compared with those from the transparent scintillator (red), which serves as a control sample. To allow relative changes in light collection between the three fibres and the PMT to be seen easily, the axes of each bar chart were scaled so that the red bars are all the same height (grey dashed line).
    The high opacity data show a clear increase (around 2.0$\times$) at the bottom of the vessel and a decrease (around 0.5$\times$) at the top, as expected from stochastic light confinement around the energy deposition point.
    Given that the low and high opacity samples have the same composition and differ only in temperature, 
    these results show that the formation of a light-ball and the corresponding increase in light collection at the bottom fibre are directly linked to the shorter scattering length.
	}
	\label{Fig_3}
\end{figure*}

An experimental proof of principle has been successfully completed with a small detector prototype. 
The setup focused on demonstrating the primary feature of our technique, which is the stochastic confinement of light. The test was done with point-like $e^-$ energy depositions to demonstrate the formation of the characteristic light-ball. 

Well-established technological solutions were used in the prototype for both the readout and the scintillator base. The latter was formulated from transparent LAB with a PPO wavelength-shifter at 2~g~l$^{-1}$. 
The opacity was obtained by mixing in a paraffin polymer at 10\% to give a uniform, waxy consistency~\cite{Ref_NOWASH}. 
Like in many waxes, the resulting scintillator was observed to transition from a transparent liquid phase at \textgreater30$\,\degree$C to an opaque white solid phase at \textless15\,$\degree$C.
This temperature dependence was exploited in the demonstration, as explained below.
The scintillator was poured into a prototype detector that consisted of a small (0.25~litre and 5.0~cm height) cylindrical vessel with internally reflecting surfaces.
Three identical Kuraray B-3 wavelength-shifting fibres were run along diametrical lines at different heights, as shown schematically in Fig.~\ref{Fig_3}a, and read out with Hamamatsu S12572-050 SiPMs. 
The detector was exposed to a mono-energetic 1~MeV $e^-$ source~\cite{Ref_BeamSource} impinging from the bottom through a thin 25~$\micro$m aluminised Kapton sheet. The $e^-$'s deposited their energy in the first few millimetres of scintillator. 

The results from the prototype are shown in Fig.~\ref{Fig_3}b. 
Three scintillator configurations were utilised: transparent (no added polymer), low opacity, and high opacity.
The former was a control sample and the latter two were obtained by setting the temperature of the same sample of opaque scintillator to 26$\degree$C and 10$\degree$C respectively. 
We note that studies of an LAB-based scintillator showed only percent-level effects on the light yield from a similar temperature change~\cite{Sorensen:2018skx}.
Direct comparison of the relative fibre response between the transparent and opaque scintillators allowed common systematic uncertainties to cancel, making the use of simulations unnecessary.
In the transparent case, the PMT saw the most light and the fibres saw different light levels consistent with their respective solid angle acceptance as the dominant effect.
When the opaque scintillator was used the light seen by the PMT and the top fibre 
was predictably reduced by a large factor. This light was not simply lost. 
The remarkable increase in light collection by the middle and bottom fibres ruled out an absorption-only scenario and showed that the light was stochastically confined around the point-like energy deposition at the bottom.
The measurements at different heights sampled the longitudinal profile of the corresponding light-ball, confirming the LiquidO detection principle.

An interesting byproduct of this measurement was the observation of temperature controlled solidification of the waxy material. 
This could open the door to several possibilities, such as doping scenarios not bound by chemical stability constraints.
The solidification also grants additional mechanical support for the fibre lattice and protection against leaks.

\subsection*{\label{sec:nuphys}Neutrino Physics with LiquidO}

\noindent The LiquidO approach is likely to open up opportunities in neutrino research. Here, we highlight a few measurements at the MeV scale where LiquidO could have a significant impact. This energy range alone provides a rich landscape of challenging physics with a wide potential for discovery. 

The performance of a LiquidO detector in terms of its position, timing and energy resolution as well as its light level and particle identification capability, depends on configurable parameters such as fibre pitch and scintillator formulation (scattering length, light yield). These parameters must be optimised for each experimental scenario by balancing all the factors at play, from the physics case to site-specific constraints (shielding, overburden) and even cost limitations. 
Prospects for specific detector implementations in concrete experimental scenarios will be studied in subsequent publications.

\paragraph*{Physics Potential with Antineutrinos.}

Above 1.8~MeV $\bar\nu_e$'s can undergo an IBD interaction resulting in a prompt $e^+$ signal followed by a delayed $n$ capture as the observable.
This is the primary channel to detect $\bar{\nu}_e$'s emitted by nuclear reactors~\cite{Qian_2019}, supernovae~\cite{Ref_SNNu}, and the earth~\cite{Ref_EarthNu}, as well as to search for these particles in decay-at-rest beams~\cite{osti_1419026,lsnd1996,karmen1999}.

In current LSDs, single $e^+$ events are largely indistinguishable from naturally occurring $e^-$'s and $\gamma$'s  with the same visible energy. Neutron backgrounds, originating mainly from the nuclear interactions initiated by cosmic-ray muons, are largely unavoidable. Furthermore, a correlated, prompt and point-like energy deposition can precede the capture of some of these neutrons and mimic a $e^+$ in an LSD. The unique signature of a $e^+$ event in a LiquidO detector, as shown in Fig.~\ref{Fig_1}c, provides a powerful handle to reject some of these backgrounds. 
As shown in Fig.~\ref{Fig_2}c, $e^-$'s or other particles giving point-like energy deposition are estimated to be misidentified as a $\gamma$ with probability of at most $10^{-2}$, which gives a reasonable estimate on the probability of misidentifying them as a $e^+$. Thus, all backgrounds whose prompt-like signals consist of $e^-$, $\alpha$ or recoil-$p$ can be reduced by a factor of at least a hundred in comparison to the latest LSDs~\cite{Ref_DCIV}.
This includes all correlated backgrounds of cosmogenic origin, which typically bear the largest impact on the background systematic uncertainty, as well as the accidental backgrounds involving a $\beta^-$.
Any remaining accidental backgrounds dominated by $\gamma$'s can be reduced by the spatial coincidence requirement that can be tightened by exploiting the more precise mm-scale vertex reconstruction. 
On top of those BG reductions, a decrease in the radioactivity present within the detector is possible through the elimination of the need for PMTs. As a case in point, the overall signal to background ratio of the Double Chooz near detector~\cite{Ref_DCIV}, with an overburden of barely 30~m of rock, would be reasonably expected to increase from about 20 to substantially more than 250.

IBD detection where backgrounds are hugely reduced opens the door to new explorations, such as a reactor antineutrino oscillation measurement beyond today's precision. Conversely, in scenarios where a high signal-to-background ratio is unnecessary, a major reduction of the overburden and shielding requirements would be possible.
A promising application in this situation would be to monitor reactor antineutrinos for non-proliferation purposes~\cite{Ref_NuReactorMonitor}.
The opacity of our technique also allows for a more efficient use of precious laboratory space. In LSDs, a passive ``buffer'' region is often used to shield against PMT or other radioactivity, and has been up to about 4$\times$ larger in volume~\cite{Ref_DCIV} than the neutrino target in past experiments. This buffer is necessary for a monolithic detector since light from energy deposited anywhere in the central region can in principle be seen by all the photo-sensors. In contrast, a segmented detector effectively keeps events localised and can thus typically tolerate higher rates. The self-segmentation of the LiquidO approach means a buffer is no longer a crucial design feature and can be greatly reduced or removed. 

\begin{figure}[t!]
	\centering
	\includegraphics[scale=0.46]{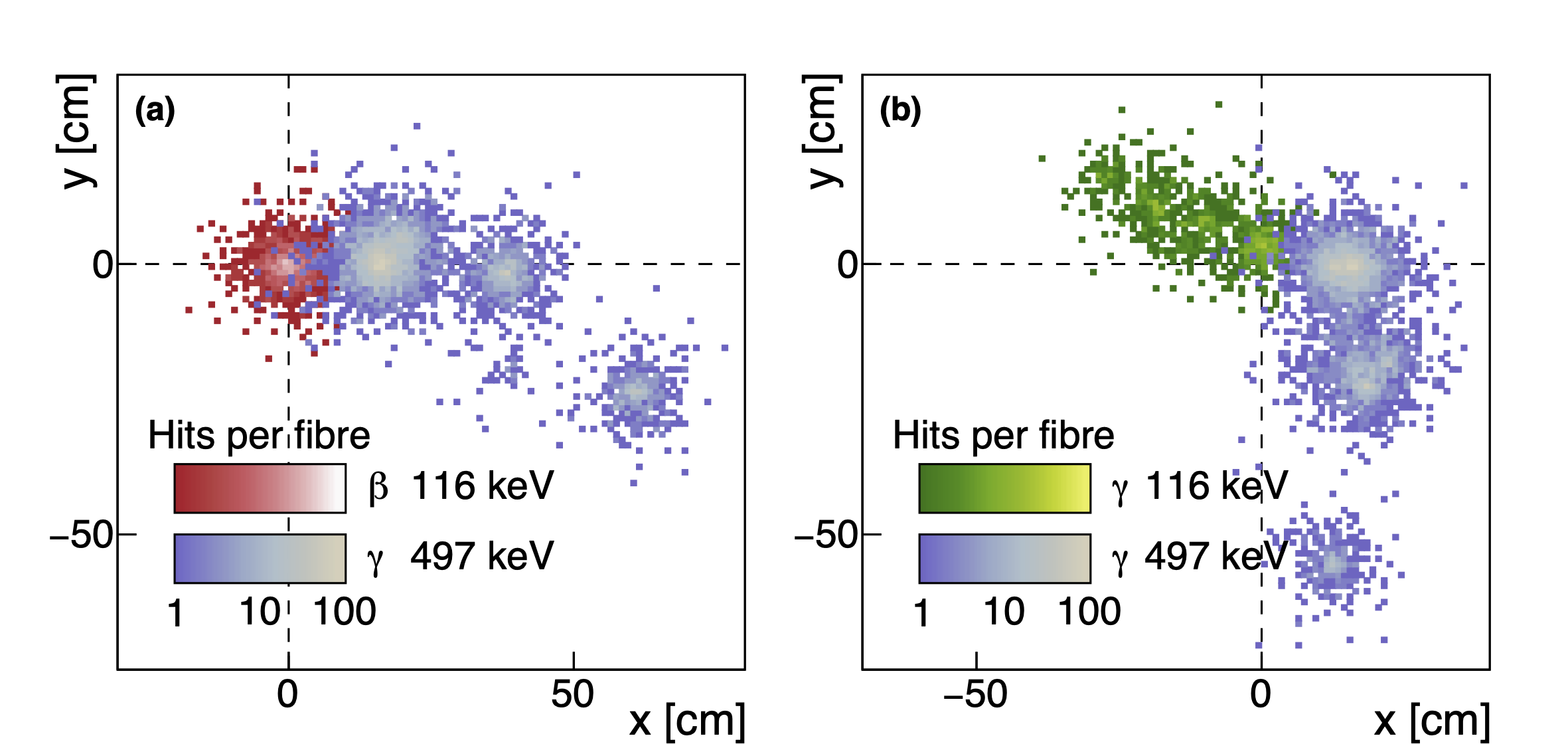}
	\caption{ \small 
	{\bf Combining LiquidO with $^{115}$In to Enable Low-Threshold Electron Neutrino Measurements.}
	At an energy of greater than 114~keV electron neutrinos undergo a charged current (CC) interaction on $^{115}$In, which could be added as a dopant at high concentrations to a LiquidO detector. 
	The $e^-$ produced has an energy proportional to the original neutrino and the excited $^{115}\mbox{Sn*}$ nucleus decays in delayed-coincidence with $\tau=4.76$~$\micro$s. 
	The two plots show a LiquidO simulation of the number of photons hitting each fibre in a 1-cm-pitch lattice for the two possible decay modes of $^{115}\mbox{Sn*}$. 
	In both cases a 497~keV $\gamma$ is produced together with either an e$^-$ ({\bf a}) or a $\gamma$ ({\bf b}) at 116~keV. 
	The colour of each point in the images represents the number of photons hitting a fibre at that $x$-$y$ location. 
	The energy deposit of the prompt $e^-$ is not shown in the plots but provides the time and space coordinates to look for the mono-energetic tin decay in delayed-coincidence. 
	Furthermore, the high-resolution images enable efficient background rejection by requiring a point-like energy deposition of the $e^-$ to be followed by the spatially dispersed Compton-scattering pattern of a $\gamma$. 
	This distinct signature that arises from combining indium with a LiquidO detector exemplifies the potential of the technique.
    }
	\label{Fig_4}
\end{figure}

\paragraph*{Physics Potential with Neutrinos.}
The detection of MeV-scale $\nu_e$ is in general a much greater challenge than $\bar\nu_e$ in LSDs. A $\nu_e$~CC interaction produces an $e^-$ in the same way a $\bar\nu_e$ produces a $e^+$, but typically without an accompanying neutron. Measuring those single $e^-$'s in LSDs is extremely hard due to the indistinguishable $\beta$'s and $\gamma$'s from natural radioactivity. It has, however, been done by experiments that went to enormous efforts to improve the radio-purity of their detection volumes~\cite{Ref_BX}. With LiquidO the dominant gamma backgrounds could be largely rejected by exploiting the difference in event topology.

An additional factor is that pure organic scintillators provide no $\nu_e$~CC interaction below 15~MeV with a high-enough yield to be useful, except for elastic scattering with $e^-$'s. The ability to dope a LiquidO detector with various elements at concentrations that would be prohibitive in conventional LSDs could enable measurements of electron neutrinos from a variety of sources that include the sun~\cite{Ref_SunNu}, supernovae and decay-at-rest beams.

In 1976, the possibility of doping with indium to enable MeV-scale $\nu_e$~CC interactions in a detector was proposed by Raghavan~\cite{Ref_Indium}. 
The interaction is $\nu_e + \,^{115}\mbox{In} \to e^- + \,^{115}\mbox{Sn*}$ and has a threshold of only 114~keV. 
The energy of the $e^-$ is proportional to that of the incoming $\nu_e$ and the excited tin nucleus decays in delayed coincidence with $\tau=4.76$~$\micro$s. 
The tin decay produces a 497~keV $\gamma$ along with either a $\gamma$ or $e^-$ at 116~keV. 
Simulated images of these two tin decays in a LiquidO detector are shown in Fig.~\ref{Fig_4}. 
Indium is 95.7\% $^{115}$In and it has been shown that stable scintillators with up to 10\% indium by weight can be achieved~\cite{Ref_DopingLENS_1, Ref_DopingLENS_2}.
The signature of the $\nu_e$ interaction on indium in a LiquidO detector is powerfully distinct. 
The prompt $e^-$ provides the time and space coordinates to look for the mono-energetic tin decay in delayed coincidence. 
Furthermore, the high-resolution images enable efficient background rejection by requiring a point-like energy deposition of the $e^-$ to be followed by the spatially dispersed Compton-scattering pattern of a $\gamma$. 
Should the rejection power be large enough to address the intrinsic $^{115}$In $\beta^-$ decay ($Q=497$~keV, $\tau=6\times10^{14}$~years) then even neutrinos from the main solar fusion chain (99.5\% of the flux) might be within reach, as was originally envisaged.

Similarly, using $^{208}$Pb as a dopant could create new opportunities. In addition to significantly enhancing the $\nu_e$ CC cross-section, particularly at higher energies (tens of MeV), $^{208}$Pb would boost the detection of the total flux from all species of neutrinos from neutral current interactions compared to what can be done with the $^{12}$C naturally present in organic scintillators~\cite{Ref_Pb208,Ref_C12}. These capabilities would enable important measurements, such as the extraction of spectral information for the high-temperature neutrinos of a supernova burst~\cite{Ref_Pb208}. Furthermore, the simultaneous detection and identification of $\nu_e$ and $\bar\nu_e$ events could enable a measurement of leptonic charge conjugation parity symmetry violation~\cite{Ref_CPVV} or other sub-dominant transitions using a pion decay-at-rest beam, where a mixture of primarily three types of neutrinos, $\nu_\mu,\,\bar\nu_\mu\:~{\rm and}\:~\nu_e$, is produced. 

Many other dopants can be considered for a range of physics purposes. Exploratory studies optimising the detector parameters and balancing the physics goals with any drawbacks introduced by doping must be carried out. For instance, the ability of a LiquidO detector to discriminate particles with point-like energy depositions from $\gamma$'s, as shown in Fig.~\ref{Fig_2}, is intrinsically reduced in the presence of heavy metals. Studies looking at specific scenarios are ongoing and will be presented in future publications.

\subsection*{\label{sec:discussion}Conclusions}

\noindent Our detector technique builds upon decades of existing expertise using scintillator detectors, but departs from the ubiquitous transparency-based approach by exploiting an opaque scintillator medium. The result is a detector that preserves many advantages of conventional liquid scintillator detectors while adding detailed imaging of particle interaction topology that enables individual events to be identified. 
With this powerful background rejection capability and the possibility of loading suitable dopants at high concentrations, a wide panorama of opportunities becomes available in MeV-scale neutrino physics and beyond. Current work focuses on further exploring the physics capabilities while continuing hardware R\&D towards larger detectors. 

\section*{Methods}
{\small
The key details of the simulation used for this paper are as follows. We used Geant4 version 4.10.04 \cite{AGOSTINELLI2003250,1610988,ALLISON2016186} to produce lists of energy deposits from particle interactions in our detector geometry. In the next step, 10,000 scintillation photons per MeV were generated within Geant4 and propagated through the geometry. 
A simple detector geometry with a 1~cm pitch lattice of 0.5~mm diameter fibres running along the $z$-axis was used, unless otherwise stated. The behaviour of the photons in the opaque scintillator was modelled using a scattering length ($\lambda_s$) and an absorption length ($\lambda_a$). We used $\lambda_s=5$~mm and $\lambda_a=5$~m, unless otherwise stated. With the 1~cm lattice, 90\% of the photons scatter until they hit a fibre with the remaining 10\% being absorbed in the opaque scintillator. The fraction of light hitting the fibres is a weak function of the scattering length: with $\lambda_s$ between 1~mm and 100~mm the collection efficiency changes by only a few percent. Additionally, any edge effects due to photons reaching the detector sides are negligible: 90\% of the photons hit a fibre within a 15~cm diameter cylinder about the light production point, and a detector several metres across was modelled. The probability of a scintillation photon being absorbed by the wavelength-shifting dye in the core of the fibre was modelled using an absorption length of 0.7~mm. The cladding of the fibre is simulated as 2 layers, each comprising 3\% of the total fibre radius and with an effectively negligible absorption length of 16~m. The outer cladding has a refractive index $n=1.42$ and the inner cladding has $n=1.49$.

The fibres were modelled as having a maximum wavelength-shifting efficiency of 90\% and a trapping efficiency of 10\% for the re-emitted photons. Both ends of the fibres were assumed to be read out using SiPMs with a 50\% photon detection efficiency. For the photons that hit a fibre, this resulted in a probability of detection of 4.5\%. Since 90\% of scintillator light hits a fibre, the overall efficiency was 4.05\% before attenuation of the fibres to their own light ($\lambda=5$~m) was applied, corresponding to a total 405 detected photons per MeV. In some studies, where stated, 300 detected photons per MeV was used, corresponding to an overall efficiency of 3\%. This included the attenuation from a 1.5~m average distance travelled in the fibres, corresponding to a 3~m tall detector.

}

\section*{Data availability}
{\small
The data supporting the findings of this study are available from the corresponding author on reasonable request.
}

\section*{Code availability}
The code that supports the findings of this study is available from the corresponding author upon reasonable request.

{\small
\bibliographystyle{unsrt}
\bibliography{bibliography}

\begin{thebibliography}{10}

\bibitem{Ref_NuDiscovery}
C.~L. Cowan, F.~Reines, F.~B. Harrison, H.~W. Kruse, and A.~D. McGuire.
\newblock {Detection of the free neutrino: A Confirmation}.
\newblock {\em Science}, 124:103--104, 1956.

\bibitem{Ref_LS}
J.~B. Birks.
\newblock {\em The Theory and Practice of Scintillation Counting}.
\newblock International Series of Monographs in Electronics and
  Instrumentation. Pergamon, 1964.

\bibitem{Ref_PMT}
W.~R. Leo.
\newblock {\em {Techniques for Nuclear and Particle Physics Experiments}}.
\newblock {Springer-Verlag Berlin Heidelberg}, 2 edition, 1994.

\bibitem{Ref_JUNO}
F.~P. An et~al.
\newblock {Neutrino Physics with JUNO}.
\newblock {\em J. Phys.}, G43(3):030401, 2016.

\bibitem{Ref_DopingGeneral}
C.~Buck and M.~Yeh.
\newblock {Metal-loaded organic scintillators for neutrino physics}.
\newblock {\em J. Phys.}, G43(9):093001, 2016.

\bibitem{Ref_RC-Loading}
A.R. Ronzio, L.~Cowan~Jr., and F.~Reines.
\newblock {Liquid Scintillators for Free Neutrino Detection}.
\newblock {\em Review of Scientific Instruments}, 29:146–147, 1958.

\bibitem{Ref_NOvA_1}
D.~S. Ayres et~al.
\newblock {The NOvA Technical Design Report}.
\newblock {FERMILAB-DESIGN-2007-01}, 2007.

\bibitem{Ref_NOvA_2}
M.~A. Acero et~al.
\newblock {New constraints on oscillation parameters from $\nu_e$ appearance
  and $\nu_\mu$ disappearance in the NOvA experiment}.
\newblock {\em Phys. Rev.}, D98:032012, 2018.

\bibitem{Ref_PDGnu}
M.~Tanabashi et~al.
\newblock {Review of Particle Physics}.
\newblock {\em Phys. Rev.}, D98(3):030001, 2018.

\bibitem{Ref_SiPM_1}
P.~Buzhan et~al.
\newblock {Silicon photomultiplier and its possible applications}.
\newblock {\em Nucl. Instrum. Meth.}, A504:48--52, 2003.

\bibitem{Ref_SiPM_2}
D.~Renker.
\newblock {Geiger-mode avalanche photodiodes, history, properties and
  problems}.
\newblock {\em Nucl. Instrum. Meth.}, A567:48--56, 2006.

\bibitem{Ref_TimePerSiPM_LAL}
V.~Puill et~al.
\newblock {Single photoelectron timing resolution of SiPM as a function of the
  bias voltage, the wavelength and the temperature}.
\newblock {\em Nucl. Instrum. Meth.}, A695:354 -- 358, 2012.

\bibitem{Ref_BX}
C.~Arpesella et~al.
\newblock {Measurements of extremely low radioactivity levels in BOREXINO}.
\newblock {\em Astropart. Phys.}, 18:1--25, 2002.

\bibitem{Ref_GERDAFibre}
M.~Agostini et~al.
\newblock {Upgrade for Phase II of the Gerda experiment}.
\newblock {\em Eur. Phys. J.}, C78(5):388, 2018.

\bibitem{Ref_NOWASH}
C.~Buck, B.~Gramlich, and S.~Schoppmann.
\newblock {Novel Opaque Scintillator for Neutrino Detection}.
\newblock {\em JINST}, 14(11):P11007, 2019.

\bibitem{Ref_FibreAcceptance}
J.~D. Weiss.
\newblock Trapping efficiency of fluorescent optical fibers.
\newblock {\em Optical Engineering}, 54(2):1 -- 5, 2015.

\bibitem{Ref_LAB}
M.~C. Chen.
\newblock {The SNO liquid scintillator project}.
\newblock {\em Nucl. Phys. Proc. Suppl.}, 145:65--68, 2005.

\bibitem{Ref_DopingLENS_1}
N.~A. Danilov et~al.
\newblock {A Study of Indium Extraction with Carboxylic Acids with the Aim to
  Produce Scintillators for Solar Neutrino Detection by LENS Spectroscopy of
  Low-Energy Neutrino}.
\newblock {\em Radiochem.}, 47(5):487--493, 2005.

\bibitem{Ref_DopingLENS_2}
C.~Grieb, J.~Link, and R.~S. Raghavan.
\newblock {Probing active to sterile neutrino oscillations in the LENS
  detector}.
\newblock {\em Phys. Rev.}, D75:093006, 2007.

\bibitem{Ref_DopingLENS_3}
C.~Buck et~al.
\newblock {Luminescent properties of a new In based organic liquid
  scintillation system}.
\newblock {\em J. Lumin}, 106:57--67, 2004.

\bibitem{Ref_KLZ}
A.~Gando et~al.
\newblock {Search for Majorana Neutrinos near the Inverted Mass Hierarchy
  Region with KamLAND-Zen}.
\newblock {\em Phys. Rev. Lett.}, 117(8):082503, 2016.
\newblock [Addendum: Phys. Rev. Lett.117,no.10,109903(2016)].

\bibitem{Ref_SNO+}
S.~Andringa et~al.
\newblock {Current Status and Future Prospects of the SNO+ Experiment}.
\newblock {\em Adv. High Energy Phys.}, 2016:6194250, 2016.

\bibitem{Ref_SNO+_2}
V.~Albanese et~al.
\newblock {The SNO+ experiment}.
\newblock {\em JINST}, 16(08):P08059, 2021.

\bibitem{Ref_JUNODBD}
J.~Zhao, L.~J. Wen, Y.~F. Wang, and J.~Cao.
\newblock {Physics potential of searching for $0\nu\beta\beta$ decays in JUNO}.
\newblock {\em Chin. Phys.}, C41(5):053001, 2017.

\bibitem{Ref_Talk_NOW}
A.~Cabrera.
\newblock {LiquidO: First Opaque Detector for ${\beta\beta}$ Decay?}
\newblock {\em PoS}, NOW2018:028, 2019.

\bibitem{Ref_BeamSource}
C.~Marquet et~al.
\newblock {High energy resolution electron beam spectrometer in the MeV range}.
\newblock {\em JINST}, 10(09):P09008, 2015.

\bibitem{Sorensen:2018skx}
A.~Sörensen et~al.
\newblock {Temperature quenching in LAB based liquid scintillator}.
\newblock {\em Eur. Phys. J.}, C78(1):9, 2018.

\bibitem{Qian_2019}
X.~Qian and J.C. Peng.
\newblock Physics with reactor neutrinos.
\newblock {\em Reports on Progress in Physics}, 82(3):036201, 2019.

\bibitem{Ref_SNNu}
K.~Scholberg.
\newblock {Supernova Neutrino Detection}.
\newblock {\em Ann. Rev. Nucl. Part. Sci.}, 62:81--103, 2012.

\bibitem{Ref_EarthNu}
O.~Sramek, W.~F. McDonough, and J.~G. Learned.
\newblock {Geoneutrinos}.
\newblock {\em Adv. High Energy Phys.}, 2012:235686, 2012.

\bibitem{osti_1419026}
A.~Bolozdynya et~al.
\newblock {Opportunities for Neutrino Physics at the Spallation Neutron Source:
  A White Paper}.
\newblock Available at https://www.osti.gov/biblio/1419026, 2012.

\bibitem{lsnd1996}
C.~Athanassopoulos et~al.
\newblock {Evidence for
  ${\overline{\ensuremath{\nu}}}_{\ensuremath{\mu}}\ensuremath{\rightarrow}{\overline{\ensuremath{\nu}}}_{\mathit{e}}$
  Oscillations from the LSND Experiment at the Los Alamos Meson Physics
  Facility}.
\newblock {\em Phys. Rev. Lett.}, 77:3082--3085, 1996.

\bibitem{karmen1999}
K.~Eitel et~al.
\newblock {The search for neutrino oscillations $\nu_{\mu} \rightarrow \nu_e$
  with KARMEN}.
\newblock {\em Nucl. Phys. B Proc. Suppl.}, 77(1):212--219, 1999.

\bibitem{Ref_DCIV}
H.~de~Kerret et~al.
\newblock {Double Chooz \ensuremath{\theta}$_{13}$ measurement via total
  neutron capture detection}.
\newblock {\em Nature Phys.}, 16(5):558--564, 2020.

\bibitem{Ref_NuReactorMonitor}
P.~Huber and T.~Schwetz.
\newblock {Precision spectroscopy with reactor anti-neutrinos}.
\newblock {\em Phys. Rev.}, D70:053011, 2004.

\bibitem{Ref_SunNu}
M.~Agostini et~al.
\newblock {Comprehensive measurement of $pp$-chain solar neutrinos}.
\newblock {\em Nature}, 562(7728):505--510, 2018.

\bibitem{Ref_Indium}
R.~S. Raghavan.
\newblock {Inverse beta decay of 115-In to 115-Sn*: a new possibility for
  detecting solar neutrinos from the proton-proton reaction}.
\newblock {\em Phys. Rev. Lett.}, 37:259--262, 1976.

\bibitem{Ref_Pb208}
J.~Engel, G.~C. McLaughlin, and C.~Volpe.
\newblock {What can be learned with a lead based supernova neutrino detector?}
\newblock {\em Phys. Rev.}, D67:013005, 2003.

\bibitem{Ref_C12}
C.~Volpe, N.~Auerbach, G.~Colo, T.~Suzuki, and N.~Van~Giai.
\newblock {Microscopic theories of neutrino C-12 reactions}.
\newblock {\em Phys. Rev.}, C62:015501, 2000.

\bibitem{Ref_CPVV}
M.~Grassi, F.~Pessina, A.~Cabrera, S.~Dusini, H.~Nunokawa, and F.~Suekane.
\newblock {Neutrino-Antineutrino Identification in a Liquid Scintillator
  Detector: Towards a Novel Decay-at-Rest-based Neutrino CPV Framework}.
\newblock {\em Nucl. Instrum. Meth.}, A936:561--562, 2019.

\bibitem{AGOSTINELLI2003250}
S.~Agostinelli et~al.
\newblock Geant4—a simulation toolkit.
\newblock {\em Nucl. Instrum. Meth.}, 506(3):250 -- 303, 2003.

\bibitem{1610988}
J.~Allison et~al.
\newblock Geant4 developments and applications.
\newblock {\em IEEE Transactions on Nuclear Science}, 53(1):270--278, 2006.

\bibitem{ALLISON2016186}
J.~Allison et~al.
\newblock Recent developments in geant4.
\newblock {\em Nucl. Instrum. Meth.}, A835:186 -- 225, 2016.

\end{thebibliography}
}

\section*{Acknowledgements}
{\small
We acknowledge the pivotal support received from the following grants:
i) the Marie Curie Research Grants Scheme (Grant 707918 between 2016-2018, fellow: Dr.~M.~Grassi hosted by Dr.~A.~Cabrera at IN2P3/CNRS) that allowed the main studies behind the simulation proof-of-principle of LiquidO;
ii) the ``Chaire Internationale de Recherche Blaise Pascal'' between 2016-2018 (Laureate: Prof.~F.~Suekane) financed by R\'egion \^{I}le-de-France (Paris, France) and coordinated by the Fondation de l'\'{E}cole Normale Sup\'erieure (Paris) and the IN2P3/CNRS via the APC Laboratory (Paris) that provided multiple levels of resources for the prototyping of LiquidO;
and
iii) the France-Japan Particle Physics Laboratory grant, since 2018, for fundamental research in particle physics cooperation between France and Japan.
We are also very thankful to 
    ANID in Chile, 
    CIEMAT in Spain,
    CNPq/FAPERJ in Brazil,
    CNRS/IN2P3 in France,
    INFN in Italy
    and 
	the University of California at Irvine in the USA
for their generous provision of manpower and resources.
%
%
We would like to acknowledge the support of the CENBG and the SuperNEMO collaboration for the use of their $e^-$ beam for the LiquidO detector prototypes.
%
%
Finally, we would like to thank several people whose knowledgeable input and kind assistance were instrumental to this publication.
These are (alphabetically):
Dr.~Y.~Lemi\`{e}re,
Prof.\,Dr.~M.~Lindner
and
Prof. Dr.~F.~Mauger. 
}

\section*{Author contributions}
{\small
All authors have contributed to this publication through their involvement in at least one of the following areas: conceiving and refining the LiquidO detection concept, developing the simulation software, designing and conducting the laboratory tests, analysing the data, and writing this article.
}

\section*{Competing interests}
{\small
The authors declare no competing interests.
}

\section*{Additional information}
{\small
Correspondence and requests for materials should be addressed to LiquidO-Contact-L@in2p3.fr.
}

\end{document}